\documentclass[12pt]{article}
\usepackage{graphicx,fancyhdr,amssymb,amsmath,geometry,lscape,float,rotating,setspace,xcolor,authblk,lineno,enumitem,wrapfig,multicol,newtxtext,gensymb,overcite,caption,subcaption,chngcntr,url,hyperref,CJKutf8,array,ragged2e}
\newcommand{\beq}{\begin{equation}}
\newcommand{\eeq}{\end{equation}}

\newcommand{\bi}{\begin{itemize}}
\newcommand{\ei}{\end{itemize}}
\newcommand{\ben}{\begin{enumerate}}
\newcommand{\een}{\end{enumerate}}

\newcolumntype{M}[1]{>{\RaggedRight\hspace{0pt}}m{#1}}
\newcolumntype{C}[1]{>{\centering\let\newline\\\arraybackslash\hspace{0pt}}m{#1}}
\usepackage{comment}
\setlist{nosep} 

\usepackage[symbol]{footmisc}

\usepackage{titlesec}
\titlespacing\section{0pt}{0pt plus 4pt minus 2pt}{0pt plus 2pt minus 2pt}
\titlespacing\subsection{0pt}{0pt plus 4pt minus 2pt}{-5pt plus 2pt minus 2pt}
\titlespacing\subsubsection{0pt}{12pt plus 4pt minus 2pt}{0pt plus 2pt minus 2pt}
\titleformat*{\section}{\Large\bfseries}
\titleformat*{\subsection}{\large\bfseries}
\titleformat*{\subsubsection}{\normalsize\bfseries}

\begin{document}

\noindent {\large{\textbf{Using satellite imagery to understand and promote sustainable development}}}

\noindent{Marshall Burke$^{1,2,3,*}$\footnote[3]{We thank Jenny Xue, Brian Lin, and Zhongyi Tang for excellent research assistance, and thank USAID Bureau for Food Security, the Global Innovation Fund, Darpa World Modelers program, and the Stanford King Center on Global Development for funding. Data and code for replication of all results will be made public upon publication. M.B., D.L., and S.E. are co-founders of AtlasAI, a company that uses machine learning to measure economic outcomes in the developing world.}, Anne Driscoll$^2$, David B. Lobell$^{1,2}$, and Stefano Ermon$^{4}$}

\noindent \footnotesize{
$^{1}$Department of Earth System Science, Stanford University, Stanford, CA \\ 
$^{2}$Center on Food Security and the Environment, Stanford University, Stanford, CA \\ 
$^{3}$National Bureau of Economic Research, Cambridge MA\\
$^{4}$Department of Computer Science, Stanford University, Stanford, CA \\ 
$^*$Corresponding author: mburke@stanford.edu, 473 Via Ortega, Stanford CA 94305.} \\ \text{ }\\


\normalsize
\textbf{Abstract} Accurate and comprehensive measurements of a range of sustainable development outcomes are fundamental inputs into both research and policy. We synthesize the growing literature that uses satellite imagery to understand these outcomes, with a focus on approaches that combine imagery with machine learning. We quantify the paucity of ground data on key human-related outcomes and the growing abundance and resolution (spatial, temporal, and spectral) of satellite imagery. We then review recent machine learning approaches to model-building in the context of scarce and noisy training data, highlighting how this noise often leads to incorrect assessment of models' predictive performance. We quantify recent model performance across multiple sustainable development domains, discuss research and policy applications, explore constraints to future progress, and highlight key research directions for the field.

\section{Introduction}

Humans have long sought to image their habitat from above the ground. Socrates purportedly stated in 500 B.C.E. that ``Man must rise above the earth -- to the top of the atmosphere and beyond -- for only thus will he fully understand the world in which he lives"\cite{moore1979picture}. His lofty goal was taken up in earnest after the advent of photography in the mid-nineteenth century C.E., with earth observation data collected by strapping cameras to balloons, kites, and pigeons.  The first known image of earth from space was taken nearly a century later (1946) by American scientists using a captured Nazi rocket, revealing blurry expanses of the American Southwest\cite{waxman2018aerial}.  This was followed decades later by the launch of the first civilian earth-observing satellite, Landsat I, in 1972, which ushered in the modern era of satellite-based remote sensing.  As of early 2020, there are an estimated 713 active non-military earth observation satellites in orbit, 75\% of which were launched in the last five years\cite{ucs2020}. These satellites are now capturing imagery of the earth in unprecedented temporal, spatial, and spectral frequency.

Here we review and synthesize a rapidly growing scientific literature that seeks to use this satellite imagery to measure and understand various human outcomes, including a range of outcomes directly linked to the Sustainable Development Goals. We pay particular attention to recent approaches that use methods from artificial intelligence to extract information from images, as these methods typically outperform earlier approaches, enabling new insights. Our focus is on settings and applications where humans themselves, or what they produce, are the outcome of interest, and where these outcomes are being predicted using satellite imagery.  We quantify existing performance in these domains across a large set of studies, explore key constraints to future progress, and highlight a number of research directions that we believe are key if these approaches are going to be improved and adopted by practitioners.  

We do not review and assess the large literature on using remote sensing for other earth observation tasks (e.g. environmental monitoring), or efforts that use other sources of non-traditional, unstructured data (e.g. data from social media or cell phones) to measure human-related outcomes. We discuss this work if these other unstructured data sources are used in combination with imagery for sustainability tasks. Our review complements existing sector specific reviews, including the use of remote sensing in agriculture\cite{mulla2013twenty,lobell2013use}, in economic applications\cite{donaldson2016view}, and in the detection of informal settlements\cite{kuffer2016slums}, drawing common lessons across these and other domains.

Our review makes four main points.  First, satellite-based performance in predicting key sustainable development outcomes is reasonably strong and appears to be improving.  Indeed, analyses suggest that reported model performance likely understates true performance in many settings, given the noisy data on which predictions are evaluated and the types of noise typically observed in sustainability applications. For multiple outcomes of interest, satellite-based estimates can now equal or exceed the accuracy of traditional approaches to outcome measurement.

Second, perhaps the largest constraint to model development is now training data rather than imagery. While imagery has become abundant, the scarcity and (in many settings) unreliability of quality ground data make both training and validation of satellite-based models difficult. Expanding the quantity and -- in particular -- the quality of labels will quickly accelerate progress in this field. 

Third, despite the growing power of satellite-based approaches, we argue that in most settings, these approaches will amplify rather than replace existing ground-based data collection efforts. Many outcomes of interest will likely never be accurately estimated with satellites; for outcomes where satellites do have predictive power, high-quality local training data can nearly always improve model performance.

Finally, there remain few documented cases where satellites have been operationalized into public-sector decision-making processes in the sustainable development domains where we focus -- with applications in population and agricultural measurements being the main exceptions. Limited adoption is likely driven by a number of forces, including the recency of the technology, the lack of accuracy (perceived or real) of the models, lack of model interpretability, and entrenched interests in maintaining the current data regime. We discuss how some of these constraints might be overcome.

\section{The availability and reliability of data}

\subsection{Key data are scarce, and often scarcest in places where most needed}

Household- or field-level surveys remain the main data collection tool for key development-related outcomes, including poverty, agricultural productivity, population, and many health outcomes. Methodologies for such data collection are well developed, and are implemented by national statistical agencies and other organizations in nearly all countries of the world. For livelihood surveys designed to generate regionally or nationally-representative estimates, sampling strategies typically follow two-stage designs, where survey ``enumeration areas" (or ``clusters", often the size of a village or a neighborhood) are first sampled proportional to population, and then a given number of households or individuals are randomly sampled within each cluster. Typically survey sizes for surveys such as the Demographic and Health Surveys (DHS) or Living Standard Measurement Surveys (LSMS) are a few hundred to a few thousand clusters, and then 10-20 households per cluster, yielding total household sample sizes typically between 2000 and 20,000 for a given country.

Such surveys provide critical information -- and often incredible detail -- on a range of outcomes, and are the bedrock on which many sustainable development related outcomes have and will continue to be measured.  But their implementation and use also faces a number of important challenges. First, nationally-representative surveys are expensive and time-consuming to conduct. Conducting a DHS or LSMS survey in one country for one year typically costs \$1.5-2 million USD\cite{sustainable2015data}, with the entire survey operation taking multiple years and involving the training and deployment of enumerators to often remote and insecure locations. Population censuses are substantially more expensive, costing tens to hundreds of millions of USD in a typical African country\cite{wardrop2018spatially}.

\begin{figure}
\begin{center}
   \includegraphics[width=\textwidth]{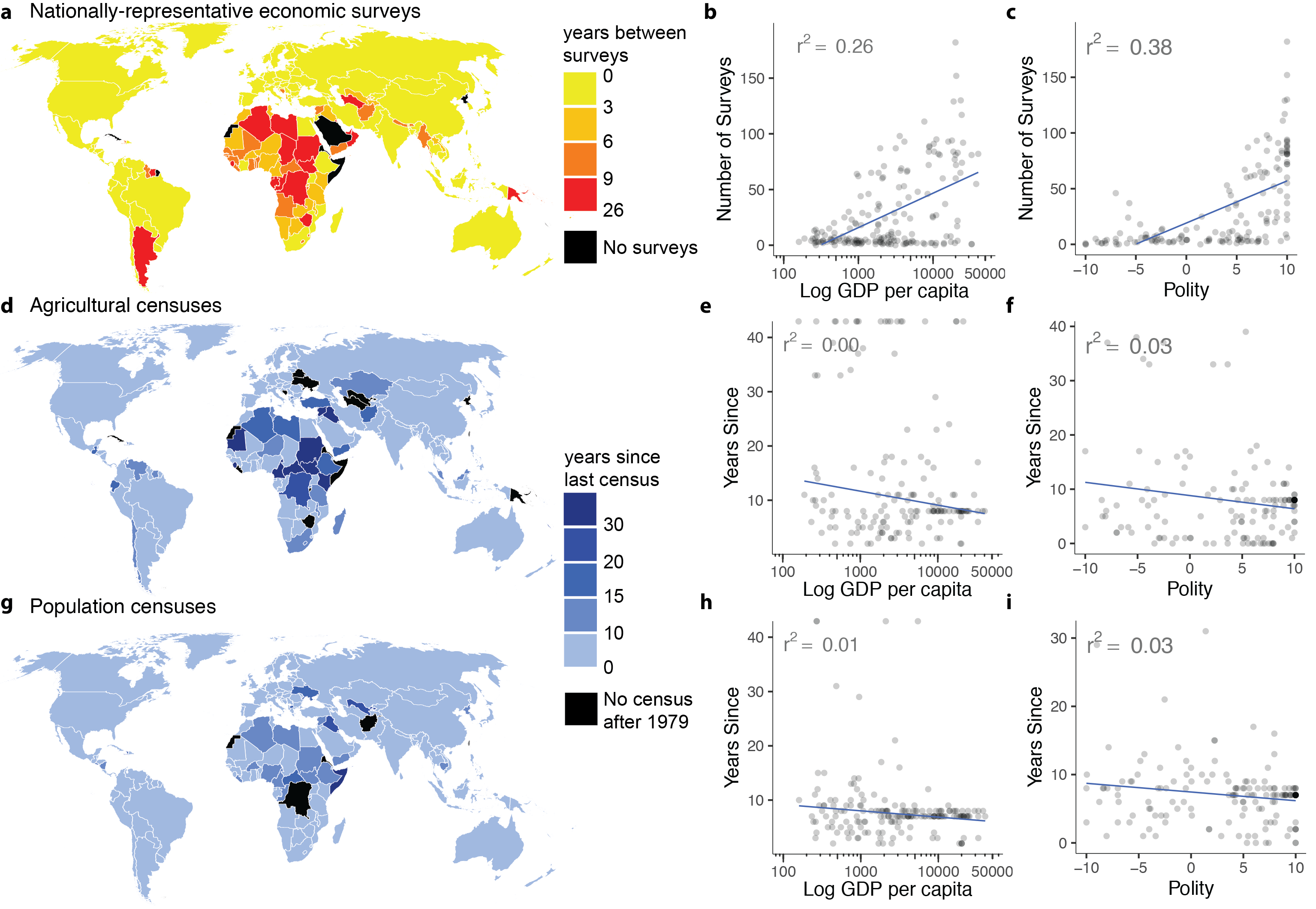}
\end{center}
\caption{\textbf{Nationally-representative economic, agricultural, and population data are collected infrequently in much of the world}.  \textbf{a} The average interval between nationally representative economic surveys (of Average or High quality) for the period of 1993 to 2018 from the UN World Income Inequality Database\cite{solt2019standardized}.  \textbf{b} Relationship between GDP per capita\cite{WB_WDI} and number of surveys in the study period. Nations with higher GDP per capita tend to have more surveys.  \textbf{c} Relationship between the Polity Score of each country (+10 is fully democratic, -10 is fully autocratic)\cite{marshall2020polity5} and the number of surveys in the study period.  \textbf{d} Years since last agricultural census. \textbf{e-f} Relationship between GDP per capita, Polity score and years since last agricultural census. \textbf{g-i} As in d-f but for population censuses.}
\label{fig:surveys}
\end{figure}

An implication of this expense is that many countries conduct surveys infrequently, if at all.  In half of African nations at least 6.5 years pass between nationally representative livelihood surveys, as shown in Figure \ref{fig:surveys}a (compare to sub-annual frequency in most wealthy countries). Globally, the frequency of these economic household surveys is on average substantially lower in less wealthy countries (Fig \ref{fig:surveys}b), meaning that data on livelihood outcomes are often lacking where they are arguably the most needed. Surveys are also much less common in less democratic societies (Fig \ref{fig:surveys}c), which could at least partly reflect the desire and ability of some autocrats to limit awareness of poor economic progress\cite{devarajan2013africa}. The frequency of agricultural and population censuses also varies widely around the world (Fig \ref{fig:surveys}d,g). For instance, 25\% (n = 53) of countries have gone more than 15 years since their last agricultural census, and 8\% (n = 17) countries more than 15 years since their last population census. Restricting to just African nations, 34\% of countries have gone more than 15 years since their last agricultural census.  For both agricultural and population data, the relationship between survey recency, income, and level of democracy is less clear, perhaps reflecting the more important role of these data in developing economies.

A second challenge for many downstream applications are that surveys are typically only representative at the national or (sometimes) regional level, meaning they often cannot be used to generate accurate summary statistics at a state, county, or more local level. This represents a challenge for a range of research or policy applications that require individual or local-level information -- for instance an anti-poverty program attempting to target an intervention (e.g cash transfer) to a particular group, or a research effort aimed at studying the impact of such an intervention.

Third, underlying household or cluster-level observations are not made publicly available in many surveys, including nearly all the surveys that contribute to official poverty statistics (such as those depicted in Fig \ref{fig:surveys}a), and no geographic information is publicly provided on where in a country the data were collected. These factors further deepen the challenge of using such data to conduct local research or policy evaluation, or to train models to predict local outcomes using these data. Even when local-level anonymized georeferenced data are made public in some form, data are typically released more than a year after survey completion, hampering real-time knowledge of livelihood conditions on the ground. 

Finally, as explored below, ground data can have multiple sources of noise or bias, further limiting their reliability and utility in research and decision-making. This in turn has important implications for how satellite-based models trained on these data are validated and interpreted.

\subsection{Existing ground data can be unreliable}
Even where ground data are present, several key sources of error can limit their utility. First, most outcomes are not measured directly, but rather inferred from responses to surveys. This can introduce large amounts of both random and systematic measurement errors, for example in the case of self-reported household consumption\cite{beegle2012methods} or agricultural production\cite{carletto2015tragedy} surveys. For instance, in household consumption expenditure surveys, changes to the recall period or the list of items households are questioned about can lead to household expenditure estimates that are $>$25\% too low relative to gold standard household diaries\cite{beegle2012methods}. 

Lack of reliability also extends to agricultural contexts. In recent reviews of agricultural statistical systems, the World Bank noted that the “practice of ‘eye observations’ or ‘desk-based estimation’ is commonly used by agricultural officers", leading to often-conflicting estimates of key agricultural outcomes by different government ministries, and to variation over time in published statistics that cannot easily be reconciled with events on the ground\cite{carletto2015tragedy}. Current practices are likely to have a bias toward overestimation, further weakening the quality of food security assessments.\cite{carletto2015tragedy,braimoh2018capacity}. 

An additional key source of noise comes from sampling variability. As noted, surveys are typically designed to be representative at very large scales (e.g. nationally), and this representativeness is typically obtained by taking small random samples of households or fields across many cluster locations.  Because most agricultural and economic outcomes of interest often exhibit substantial variation even at very local levels (e.g. coefficients of variation $>1$ at the village level), these small samples thus represent an unbiased but potentially very noisy measure of average outcomes in a given locality. %

The combined effects of both measurement error and sampling variability can be appreciated when comparing two independent measures of the same outcome for the same administrative level.  In Figure \ref{fig:agdata}, average maize yields (in units of tons per ha of land) are compared at the first administrative level (e.g., province or state) as obtained from household surveys covered by the LSMS-ISA program versus by official government ministry estimates in three African countries. This comparison reveals both a systematic bias towards higher yields in official government data than in household responses, and a relatively low correlation between the two measures, with the highest observed correlation equal to $r=0.39$ for Ethiopia. 

A third source of error, particularly relevant to researchers relying on access to data acquired by others, is noise purposefully introduced to protect the privacy of surveyed households. Adding jitter to village coordinates is common practice for most of the publicly released datasets based on household surveys, for instance with up to 2km of random jitter added in urban areas and 5km in rural areas.  Below we explore the implications of these three sources of error for model development and evaluation.

\subsection{Availability of satellite imagery changing rapidly}
Information from satellite imagery has long offered a potential inroad into helping solve problems of data scarcity and unreliability in sustainability. Such information has been used in both agricultural and socioeconomic applications for decades\cite{macdonald1980global,elvidge1997relation}.  However, thanks to both public and private sector investment, recent years have seen a remarkable increase in the temporal, spatial, and spectral information available from satellites. These investments have largely undone the traditional trade-off between temporal and spatial resolution, and are helping to undo the trade-off between spectral and temporal/spatial resolution. 

To quantify this increase and understand how it varies across developing and developed countries, we randomly sample 100 locations in Africa and 100 additional across the US and EU (sampling proportional to population), and query the availability of cloud-free imagery (defined as $<$30\% cloud cover) at each location in 2010 and 2019 for all available optical sensors, using multiple online tools (see Supplemental Information for details on this process). We calculate region- and year-specific average revisit rates as the number of available cloud-free images across locations divided by the number of locations times the number of days. We calculate this separately for each sensor and and also calculate an imagery-resolution ``frontier", defined as overall revisit rate across sensors at or below a given spatial resolution. 

Results are shown in Figure \ref{fig:imagery}. Many new public and private-sector entrants
since 2010 (Fig \ref{fig:imagery}a) have lessened the traditional temporal/spatial tradeoff in imagery, particularly at resolutions $\geq$3m. Although the revisit rate of very high resolution ($<$1m) sensors over Africa has seen only slight improvement over the last decade (Fig \ref{fig:imagery}b), and very-high-resolution revisit rates remain lower in Africa as compared to the US/EU (Fig \ref{fig:imagery}c), revisit rates for high resolution (1-5m) and moderate- to low-resolution sensors has increased dramatically.  Images at this resolution arenow captured multiple times per week rather than multiple times per year and equitable capture between Africa versus the US/EU. 

Figure \ref{fig:imagery} provides additional detail and sample imagery for a number of sensors in African locations. Information on human activity is readily visible even in moderate-resolution sensors (5-30m), and indices constructed from moderate-resolution multispectral imagery provide an increasingly clear picture of a broad range of human activity at very local scale, including urban infrastructure development, agricultural activity, and moisture availability (Fig \ref{fig:imagery}f). The increasingly high revisit rate of such imagery also provides key insight into development-relevant activities that change seasonally, such as the location and productivity of croplands (Fig \ref{fig:imagery}g).

\section{Modeling approaches using satellite imagery to predict sustainability outcomes}

Researchers have taken many different modeling approaches in using this large amount of new imagery to measure and understand sustainable development. 
We use ``model" to mean any function or set of functions mapping inputs (e.g., satellite images) to outputs (e.g., a wealth index or yield estimates for an area). 
Such models are often simple, such as linear regression models that relate satellite-derived vegetation indices to crop yields\cite{burke2017satellite} or that relate nighttime lights to economic outcomes\cite{henderson2012measuring}.
When there is substantial prior knowledge of the likely relationship between satellite-derived features and the outcome of interest, as in the case of many agricultural variables, such approaches can often work well. 
However, even in these settings, machine learning approaches that seek to more flexibly learn -- rather than specify -- the mapping of inputs to outputs can often improve predictive performance.

Machine learning approaches start by defining a suitable model family, i.e., a set of candidate functions used to represent the relationship between inputs and outputs. 
These could be decision trees, random forests, support vector machines, or fully-connected neural networks with a fixed structure and varying weights~\cite{hastie2009elements}. When inputs and outputs have explicit spatial or temporal structure (e.g. images, or images over time) it is typically advantageous to use functions tailored to this structure. These include convolutional neural networks for images, recurrent neural networks for sequential data, and convolutional autoencoders when both inputs and outputs have spatial structure~\cite{goodfellow2016deep}(e.g., segmentation of agricultural fields).

Training data for these models consists of a set of inputs with their corresponding ground-truth outputs, e.g., images of villages and their corresponding poverty levels, or a sequence of images of a field captured during the growing season and the corresponding crop yield. 
A model in the family is chosen by training, which typically involves the minimizing of a suitable loss function %
that describes the difference between predicted and observed values of the outcome. For regression, the loss could be squared loss or absolute value, and for classification a common choice is cross-entropy. 
After training,  the loss function is evaluated on held-out data -- i.e. data not used to train the model. Evaluation on held-out data is important because training data are often limited and the model family complex (often with many orders of magnitude more model parameters than training observations), and overfitting is thus a major concern. Regularization techniques such as weight decay, dropout, and early stopping using a validation set are often employed in practice to mitigate overfitting.

Suitable preprocessing of inputs is also often important in achieving good performance. Common pre-processing steps include median compositing across images to mitigate the effect of occlusions due to clouds, imputing missing values, scaling to put the all the inputs on the same scale, centering, whitening, and harmonic preprocessing for temporal data. While deep models can in principle learn these tranformations, these are tailored to existing learning algorithms and initialization schemes and will generally make learning more stable. Tiling and rescaling is also often necessary to match the input requirements (e.g. pixel dimensions) of exiting neural architectures of neural networks.

Here we provide an overview of the range of modeling approaches that have been used to relate satellite images to sustainable development outcomes. 

\textbf{Shallow models based on hand-crafted features.} In some domains, prior knowledge of the physics, chemistry, or biology of the relevant processes suggest that certain functions of the inputs are likely useful for prediction. This is the case for numerous vegetation indexes (VI), which are computed from raw imagery as simple ratios of reflectances at different wavelengths and are known to be related to vegetation health. 
Simple regression models such as linear regression or random forests can be used to make pixel-wise predictions directly from these hand-crafted features to the outputs of interest, e.g. predicting yield with VIs observed over time (see ref\cite{weiss2020remote} for a recent review in the agricultural domain). 
When the input has spatial structure, simple aggregation strategies can be used to map pixel-wise features to image-wise features. These include simple statistics such as taking the mean, quantiles (min,median,max), or histograms of binned values as inputs to a regression or ML model.
As an example, this strategy is very effective for predicting GDP with nightlights\cite{henderson2012measuring} or aggregate crop yields at the county and state level from multispectral images~\cite{you2017deep}. However, these simple aggreagation strategies discard most of the spatial structure, which can be undesirable.

\textbf{Models that use spatial structure in the imagery}.
In computer vision, spatial context can often greatly improve prediction accuracy for image prediction and analysis tasks. Machine learning models with filters designed to take into account spatial structure, such as convolutional neural networks (CNNs), often perform much better than hand-crafted features and aggreagation strategies. Models such as VGG~\cite{simonyan2014very}, or deeper models with residual connections such as DenseNET or  ResNet~\cite{he2016deep} are often employed. In this case, features are automatically learned from the data rather than hand-crafted. This is currently the leading approach in most computer vision applications, including in the satellite space when training data are plentiful. Use of this approach in sustainable development applications has proliferated in recent years, including in the measurement of population\cite{tiecke2017mapping,zong2019deepdpm,hu2019mapping}, economic livelihoods\cite{jean2016combining,head2017can,steele2017mapping,yeh2020using}, infrastructure quality\cite{cadamuro2018assigning,barak2018infrastructure}, land use\cite{albert2017using,helber2019eurosat}, informal settlements\cite{mboga2017detection,persello2017deep}, fishing activity\cite{kroodsma2018tracking,park2020illuminating}, and many others. 

\textbf{Models that use spatial and temporal structure in the imagery}.
When available, multiple images of the same location over time can reduce ambiguity (e.g., due to partial cloud cover) and provide provide crucial information about changes occurring on the ground. Such a sequence of images is similar to a video, and architectures from video prediction in computer vision can be brought to bear for prediction and regression tasks. These include recurrent neural network variants such as long-short term memory networks (LSTMs)~\cite{hochreiter1997long}, convolutional LSTMs~\cite{xingjian2015convolutional}, and 3-D CNNs, where images are fed in sequence into the model before it makes a prediction. 
These models have been successfully used for crop classification~\cite{ji20183d,russwurm2017multi,m2019semantic}, crop yield prediction~\cite{you2017deep,sun2019county}, predicting landslide susceptibility~\cite{xiao2018landslide}, assessing building damage after disasters~\cite{xu2019building,ci2019assessment} among many other tasks.
 
\textbf{Models that use several modalities}.
When multiple data modalities are available, such as measurements from different satellites, it is often possible to combine all the inputs into a single deep learning model. Approaches include stacking the inputs as additional channels of a single network, or multi-branch architectures where data modalities are processed separately to extract features which are then concatenated before a final prediction layer. Examples of this approach include models that combine multiple sources of satellite information\cite{yeh2020using} or models that combine imagery with data from weather sensors\cite{davenport2019using}, cell phones\cite{steele2017mapping}, Wikipedia\cite{sheehan2019predicting}, social media\cite{fatehkia2020relative}, street-level imagery\cite{cao2018integrating} or Open Street Map\cite{tingzon2019mapping} to predict development-related outcomes.  
 
\subsection{Model development with limited training data}

An additional set of techniques have been developed to utilize the above modeling approaches in the context of limited training data -- a common problem in sustainability applications.  For instance, standard convolutional neural network architectures contain millions to tens of millions of trainable parameters\cite{huang2017densely}, whereas training data for specific sustainability tasks can often number in the hundreds. This limited amount of labeled data is often insufficient for ``end-to-end" training of deep networks, i.e. training a model to directly predict the outcome of interest on the available labeled data by minimizing a suitable loss function.  
Multiple strategies have been deployed to address this problem. 

\textbf{Using synthetic data.}
A first approach is to generate and use synthetic data to train models. In some cases, domain knowledge about the relevant physical process exists in the form of validated simulators. These simulators can be used to provide synthetic training data, i.e., synthetic inputs of what the process would look like from space paired with simulated outputs. These synthetic pairs can be used to augment the training data. For example, crop model simulations have been used to augment field data collection for satellite-based yield mapping in smallholder systems, and have been shown to perform on par or better than approaches that calibrate directly to limited field data\cite{burke2017satellite,lobell2020eyes}. 

\textbf{Transfer learning}.
A second approach, transfer learning, is a common strategy in deep learning. The idea is that a neural network can be pre-trained on a different but related task for which large amounts of labeled data are available (such as ImageNet in computer vision, or Functional Map of the World\cite{christie2018functional} and WikiSatNet\cite{burak2019learning} for satellite images). The neural network is then ``fine-tuned" on the task of interest. For example, Jean et al\cite{jean2016combining} showed how transfer learning could use be used to predict a very small ($<$500) number of observations of economic livelihoods in Africa from imagery. A neural network was first trained to predict nightlights (a plentiful proxy for economic development) from daytime imagery, thus learning to recognize features in the high-resolution daytime imagery related to economic activity. Features were then extracted for daytime images in locations where livelihoods data were available, and a simpler model (e.g. regularized regression such as ridge or lasso) used to predict livelihoods from these features. 
Another recent approach applied a trained object identifier to high resolution data to identify buildings, vehicles, and other objects, and then used these objects as features in a regularized regression to predict economic wellbeing in Uganda with high accuracy\cite{ayush2020generating}.

Transfer learning can also be done spatially, with models trained using data from one region where labels are often plentiful, and then ``fine-tuned" on the target geography of interest where labels are sparse.  To be successful, this approach requires relevant features to be similar between training and target geographies, but does not require the mapping of features to outcomes to be the same between regions (e.g. having productive crops near your house could signify wealth in one region but relative poverty in another). 
For example, a model trained to predict infrastructure quality in Africa could be finetuned to a specific country using only a small amount of labeled data\cite{barak2018infrastructure}. The main challenge with spatial transfer learning is that changes in the input data distribution from one region to another (e.g. the appearance of houses or crops) will decrease predictive performance.

\textbf{Unsupervised or semi-supervised learning}.
A third approach uses unsupervised or semi-supervised learning, which take advantage of the fact that while labels are often scarce in sustainability applications, obtaining large amounts of unlabeled satellite imagery is relatively easy. Utilizing large amounts of unlabeled data to pre-train neural networks and learn useful features has recently shown great progress in computer vision~\cite{he2020momentum,chen2020simple}, narrowing the gap with fully supervised methods.
Among others~\cite{basu2015deepsat}, Tile2Vec is an unsupervised pre-training technique tailored specifically to satellite images that performs well on a range of tasks, such as crop type classification and predicting economic wellbeing in Africa~\cite{jean2018tile2vec}. 
Semi-supervised learning strategies attempt to improve model performance by additionally leveraging a small amount of labeled data. These are often based on the assumption that data is clustered, and decision boundaries should separate these clusters as much as possible. This idea has been extended to regression problems, with resulting performance improvements in predicting economic well-being from satellite imagery\cite{jean2018semi}.

\subsection{Model development and evaluation with noisy data}

The performance of satellite-based models, particularly in settings beyond where they were trained, is perhaps the most common and important concern for researchers and policy makers interested in potential applications in sustainable development. Noisy training data can degrade model performance in two ways. First, it can diminish the ability of a model to learn features in imagery that are predictive of the outcome of interest. Second, and more subtly, the model might learn relevant features but perform poorly in predicting test data, precisely because the test data has noise. This latter outcome would lead researchers to understate the model's true performance. As noisy datasets are increasingly employed for model development, researchers must contend with the dual challenges of not overfitting to noise and not underestimating the performance of a model with respect to reality. Both challenges are potentially important, with existing work mainly highlighting how noise in training data can degrade model performance\cite{elmes2020accounting}. But in many sustainable development settings, we believe models can learn to separate signal from noise in training data, and that the more fundamental -- and underappreciated -- challenge is in accurately assessing model performance in light of noisy test data.  We quantify this insight and discuss methods for addressing it. 

\begin{figure}
\begin{center}
   \includegraphics[width=\textwidth]{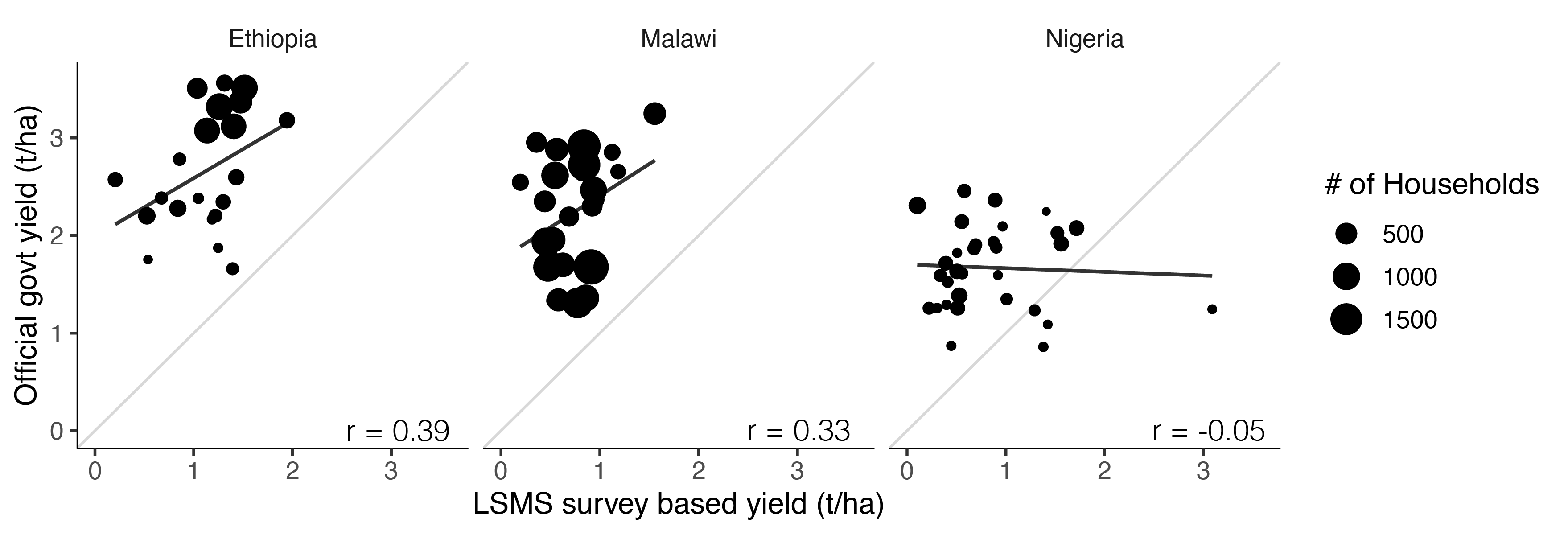}
\end{center}
\caption{\textbf{Government and household-survey based data on maize productivity are not well correlated at the district level}. Using government data from eAtlas and household level yield data from LSMS-ISA surveys, maize yields are compared by averaging across all households in a given district. Data include 2011, 2013, and 2015 data in Ethiopia, 2013 data in Malawi, and 2010 and 2012 data in Nigeria. Comparison is restricted to district-years with at least 30 households. Grey line is 1:1 line, while black lines show linear fits within each country. Points are sized relative to the number of households contributing to each estimate in the LSMS data.}
\label{fig:agdata}
\end{figure}

\textbf{Noisy training versus noisy test data}
Studies in the broader computer vision/deep learning domain have demonstrated how models trained on noisy but numerate labels can still perform well when evaluated on high-quality test data, even when high-quality labels are massively outnumbered by low-quality labels in training data\cite{krause2016unreasonable,rolnick2017deep,natarajan2013learning}.
Under suitable assumptions on the noise, these empirical results can be explained from a theoretical point of view~\cite{natarajan2013learning,charikar2017learning}. 
In sustainable development settings, while noisy training can certainly still degrade model performance when the amount of training data is limited (ref\cite{paliwal2020accuracy} provides one example in Indian smallholder wheat systems) or errors non-random (as in the poor-quality government data in Fig \ref{fig:agdata}), numerous recent studies highlight how such noise can be overcome so long as training data are reasonably numerous and errors are largely random. For instance, in Uganda, a model trained to predict maize yields from relatively noise data performed twice as well when evaluated on high-quality test data as when evaluated on noisy data held-out from training\cite{lobell2020eyes}. 
In India, a satellite-based crop classification model trained on labels derived from millions of imperfectly geolocated smartphone photos was able to exceed the performance of benchmark satellite-based classifiers\cite{wang2020mapping}. A global study showed how noisy object labels from Open Street Map could be used to train a model to make accurate predictions of the location of urban structures\cite{kaiser2017learning}.

Using data and imagery from an earlier study of asset wealth across thousands of African villages\cite{yeh2020using}, we use simulation to explore the influence on model performance of three types of error common in publicly-available training data:   (1) noise due random noise (``jitter") purposely added to village geo-coordinates to protect respondent privacy, (2) sampling variability noise from the construction of village-level estimates from small numbers of respondent households, and (3) noise from households' misreporting of asset ownership. %
We add a given type of noise to the observed wealth estimates, train a random forest model to predict these labels from nightlights imagery on 5 folds of the data, and evaluate performance on the remaining test data that has either been similarly degraded or unaltered; we use nightlights and random forest rather than a CNN and/or optical imagery to make these experiments tractable. 

\begin{figure}[H]
\begin{center}
  \includegraphics[width=\textwidth]{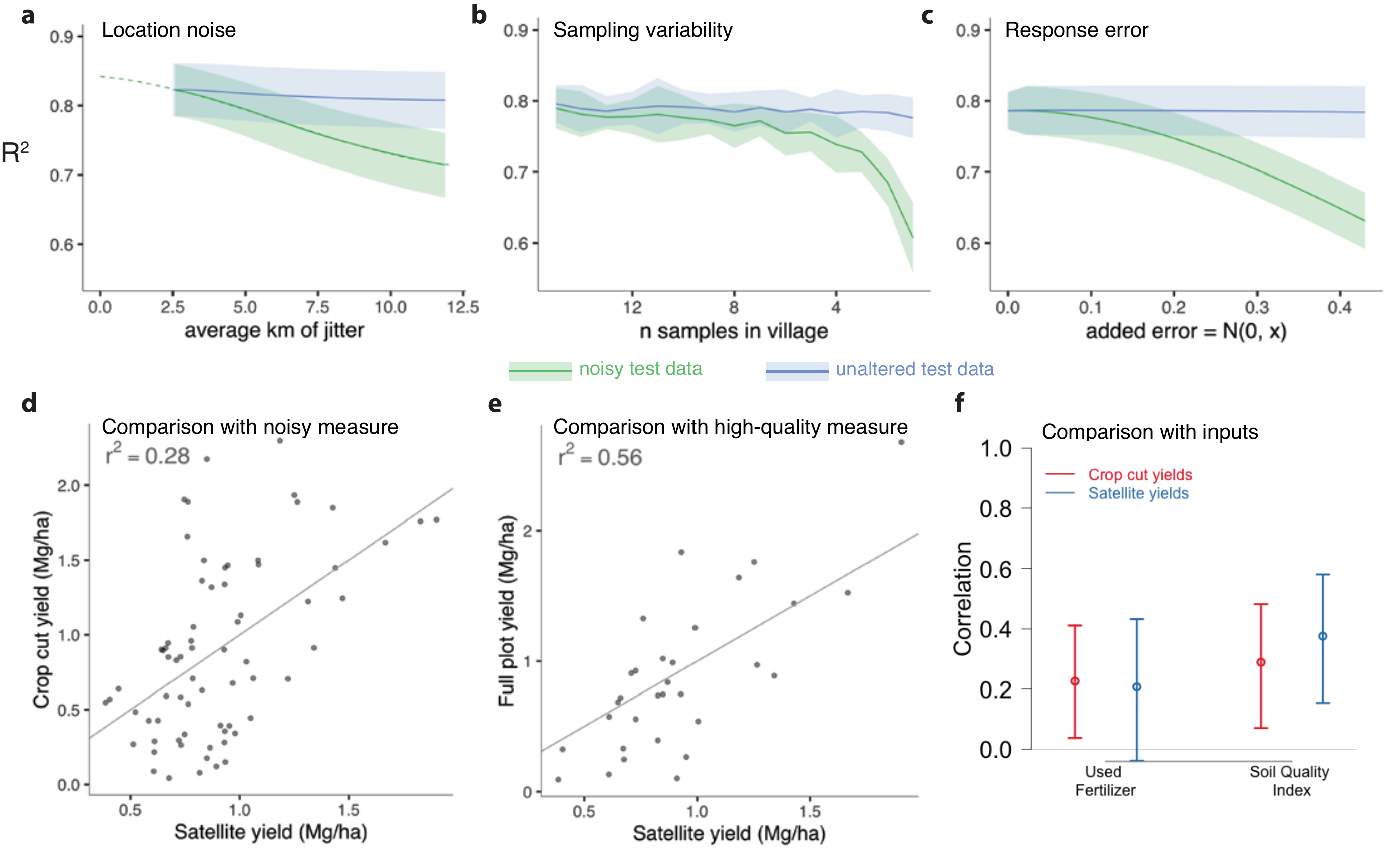}
\end{center}
\caption{\footnotesize{\textbf{The role of noise in model performance and evaluation}. \textbf{a-c} Performance of wealth prediction model as noise is added to train and test data.  Model trained to predict asset wealth from nightlights imagery across 4000 African villages, using the dataset from ref\cite{yeh2020using}. Performance is evaluated as three different types of noise are added to training data: \textbf{a} random noise in village geo-coordinates (starting from 2.5km, the actual noise in the survey data), \textbf{b} noise from constructing village-level wealth estimates from decreasing numbers of households within the village to represent sampling variability, and \textbf{c} random noise added to village-level wealth estimates, representing random response error from respondents. Green lines show performance evaluated on test data where similar noise has been added, blue lines show performance on test data where noise has not been added. Shaded areas indicate confidence intervals across 200 runs at a given level of added noise. As all types of training noise increase, model performance degrades when evaluated against similarly noisy test data but does not degrade when evaluated against unaltered test data. %
\textbf{d-f} Example from a study of maize yields in Uganda\cite{lobell2020eyes} in which both ground-based and satellite-based measurements can have noise, and multiple approaches can help adjudicate which is noisier. \textbf{d} Imperfect correlation between ground- and satellite-based yield measure does not reveal source of noise. \textbf{e} Comparison of satellite measure with available gold-standard ground measure from full plot harvest shows higher correlation, indicating ground measure in (d) responsible for at least some of the noise. \textbf{f} Comparison of satellite measure and ground measure with independent third measures expected to correlate with yields (here, fertilizer use and soil quality) suggests that the two yield measures in (d) are roughly equally noisy.}}
\label{fig:modelperformance}
\end{figure}

As shown in Fig \ref{fig:modelperformance}a-c, when evaluated on noisy training data, model performance degrades as increasing amounts of each type of noise is added. However, when models trained on increasingly noisy data are evaluated on un-degraded test data, model performance remains highly stable, even for large amounts of training noise.  This holds true for all three common types of training data noise we explore, again suggesting that ML models can be surprisingly robust to various types of training noise.

\textbf{Accurately assessing model performance}.
Most existing work has focused on techniques to avoid overstating model performance, including strategies discussed above to avoid overfitting during training, and the typical practice of testing models on held-out data. Here we discuss two strategies for dealing with the opposite problem: understating model performance due to noise in test data.

A first approach is to ensure that a small amount of very high-quality ground data is available for model testing. Often this can require additional investment in data collection.  In using these data, the typical practice of splitting a dataset from a single heritage into training, validation, and test sets is then replaced by a practice with two different measurement approaches for training and validation on the one hand, and testing on the other -- with the high-quality data reserved for testing. Typically, the data volumes needed for testing are far fewer than for training, and thus the expenses associated with obtaining “gold-standard” measures for testing are more likely tractable. 

A second strategy, particularly useful if ground data are unavailable, is to identify a variable that previous work has identified as being associated with the outcome of interest, such as weather in the case of economic output, or fertilizer in the case of agricultural productivity. This strength of association between this variable and model predictions, as measured for instance by correlation, can then be compared to the association between the variable and the (noisy) training data for the model. Because these third variables (e.g. weather) are often readily available for most locations in the world, this approach should have broad applicability.

To illustrate both of these strategies, Figure \ref{fig:modelperformance}d-f draws on a recent study of maize yields in Uganda\cite{lobell2020eyes}. The left panel shows the agreement between satellite-based yield estimates and the data on which the model was trained. In this case, the training data comprised 8mx8m crop cuts (i.e. harvests from small, randomly-selected portions of a field) from 125 different maize fields in the region. Although crop-cuts are low-error measurements of productivity for the portions of a field they sample, we consider this noisy training data because of high heterogeneity within fields and the potential spatial mismatch between the crop-cut location and the satellite pixels (which are 10x10m and not perfectly aligned with the crop cut). As judged by the training fit, the model has a relatively modest explanatory power ($r^2=0.25$, Fig \ref{fig:modelperformance}d). Yet the model performance is much better when predictions are compared to the ``gold-standard” measure of full plot harvests, which were available for a smaller number of randomly-selected fields (Fig \ref{fig:modelperformance}e).
Similarly, the correlation between satellite estimates and self-reported fertilizer or objective measures of soil quality were the same as the correlation between crop cut yields and these measures, suggesting the ``signal" in the satellite measures was as strong as that from the ground measure (Fig \ref{fig:modelperformance}f). A similar finding was obtained in Kenya when pitting satellite-estimated maize yields against self-reported yield data\cite{burke2017satellite}.

Another example of both strategies is given in ref\cite{yeh2020using}, where estimates of wealth from satellites and from ground data are each compared against independent wealth measures from census data (considered high quality) and against a measure of annual temperature, which has been shown to correlate strongly to economic outcomes. Ground data and model predictions showed similar correlation against the independent wealth measure, and both uncovered similar non-linear relationships between temperature and wealth, suggesting that the satellite-based wealth measure was roughly as trustworthy as the original ground data.

\section{Applications}

Researchers are actively evaluating the usefulness of satellite imagery for a range of sustainable development applications, with more work thus far focused on whether satellites can be used to make reliable measurements of key variables of interest and comparatively less devoted to using derived measures for downstream research tasks or policy decisions.  Rather than try to provide a comprehensive survey of all applications of satellite-based remote sensing in sustainable development, we focus on four domains where recent work on satellite-based measurement has been particularly active and where comparable quantitative results exist across studies. Our goal is to provide rough performance benchmarks across these domains and, where possible, diagnose constraints to further improvement. In making these comparisons, we included all published or posted (e.g. on arxiv) studies where the test statistic of interest could be obtained for the outcome of interest in a developing-world geography.

We then review the more limited set of cases where these and other satellite-based measurements have been used for research or policy tasks. Our focus is again on domains directly involving human activity, and does not encompass progress in all realms of earth or environmental observation.

\textbf{Smallholder agriculture}.
Roughly 2.5 billion individuals, and over half of the worlds poor, are estimated to live in ``smallholder" households that primarily depend on farming small plots of land for their livelihoods\cite{christen2013segmentation}. While remote sensing has been used in agricultural applications for decades, coarse sensor resolutions and a paucity of training data had until recently largely precluded its application in smallholder agriculture, where field sizes are often $<$0.1ha (or roughly 1 30m Landsat pixel). 

\begin{figure}[H]
\begin{center}
   \includegraphics[width=0.9\textwidth]{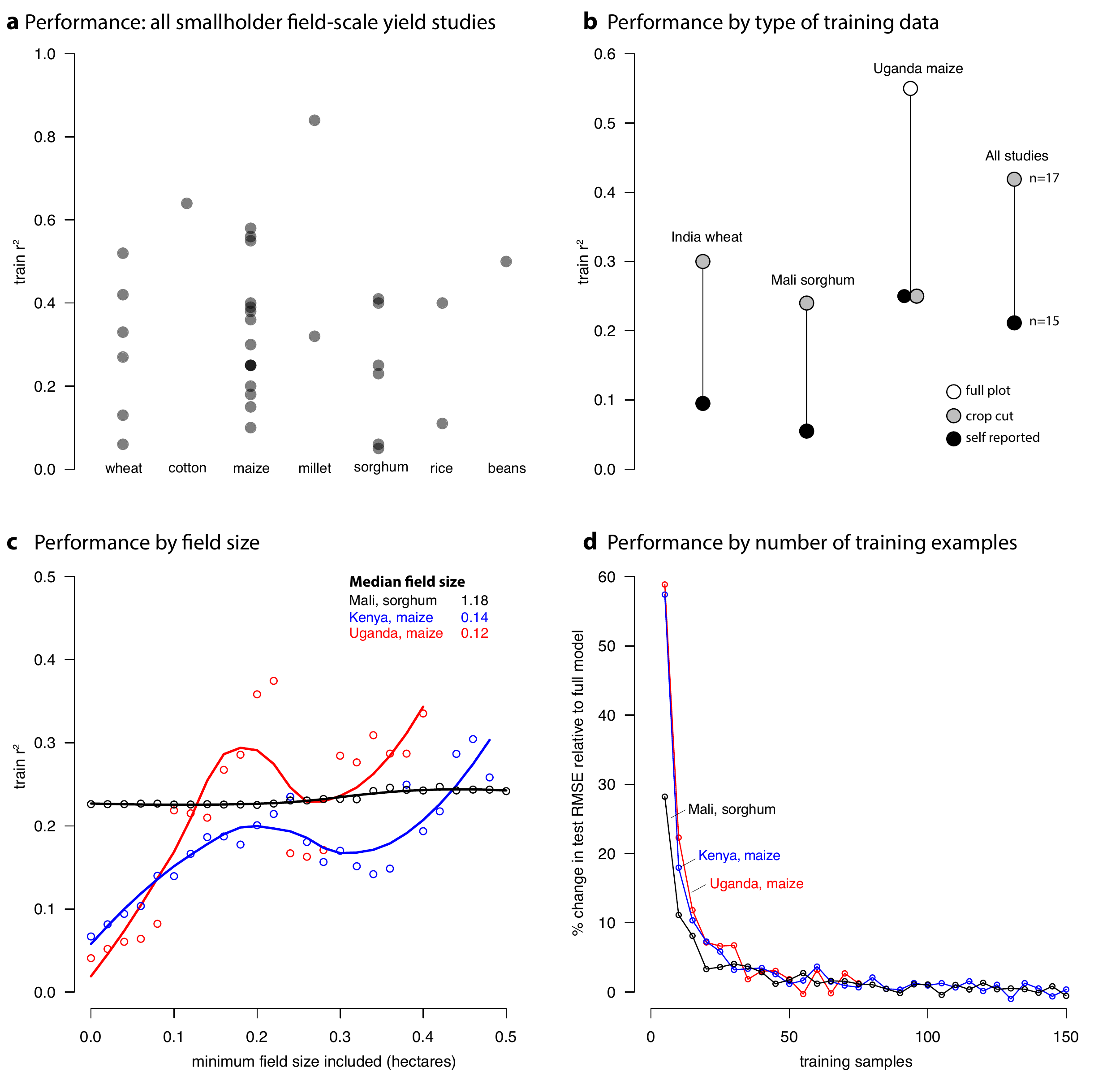}
\end{center}
\caption{\textbf{Performance of satellite-based approaches to measuring smallholder yield at field scale}. \textbf{a} Performance across all known published studies where coefficient of determination ($r^2$) was reported (32 estimates across 11 studies); $r^2$ estimates are ``in-sample", i.e. for data on which model was trained. \textbf{b} Difference in performance for models trained and evaluated on crop-cut, self-reported, or full-plot harvest data suggest that more objective crop measures improve performance. First three estimates are for studies that compared at least two types of ground data in the same setting. ``All studies" estimates pool across estimates in (a). \textbf{c} Performance generally increases when sample is restricted to larger fields, particularly in East African settings where field sizes are very small. \textbf{d} Performance on test data improves rapidly with additional training examples up to $\sim$30 data points, and then improves more gradually thereafter. Performance measured as average root mean squared error between predicted and observed yields in the test set, averaged over 100 different random subsets of training samples at each size of the training set.}
\label{fig:smallholder}
\end{figure}

Here we assemble data from recent studies attempting to predict yield at the field scale in heterogeneous smallholder environments (ref\cite{chivasa2017application} provide a nice overview of yield prediction performance at more aggregate scales). Field-scale yield prediction is useful for a range of development applications, including the targeting and evaluation of agricultural interventions and the rapid monitoring of rural livelihoods. We found 11 published studies that reported comparable performance metrics for field-scale yield prediction on smallholder fields, spanning multiple continents and seven crops. All studies used relatively simple models to relate hand-crafted features (typically, vegetation indices constructed from ratios of reflectances in the visible and near-infrared wavelengths) to ground-measured yields, and nearly all evaluated models on training rather than held-out test data.  While predictive performance differed widely across and within crops (Fig \ref{fig:smallholder}a), likely due to the enormous temporal and spatial heterogeneity present in smallholder agriculture, re-analysis of multiple studies for which replication data were available allowed insight into the determinants of model performance. 

First, models trained and evaluated on more ``objective" ground data -- i.e. harvest data collected from crop cuts or full plot harvests -- performed on average substantially better than models trained on farmer self-reported data (Fig \ref{fig:smallholder}b). This finding again highlights the importance of ground-based measurement error in training and evaluating remote sensing models. Second, in settings where average field sizes were small, model performance was much higher on larger fields (Fig \ref{fig:smallholder}c). This difference is likely because for certain sources of error, e.g. error in field area measurement or in the georeferencing of field data, the same magnitude error is more consequential for smaller fields; a 10m georeferencing error is more consequential for a 10m-wide field as compared to a 100m-wide field. %

Finally, because collecting high quality ground data is expensive and time consuming, we studied the extent to which additional training samples improve model performance.  At very small sample sizes, additional training samples rapidly improved performance on held out test data (as measured by root mean squared error; Fig \ref{fig:smallholder}d), up to around 30-50 samples. Performance was largely stable beyond that, suggesting that -- at least in the African settings represented here -- adequate performance for yield prediction could be achieved with only a few dozen high quality training samples. See Table \ref{tab:ag} for the full list of studies and estimates we included.

\textbf{Population} A second area in which satellite information has played an important role is helping generate local-level population estimates. Accurate knowledge of where people are is a critical input into an immense range of research and policy applications.  Because population census are infrequent in many developing countries and fine-scale data from existing censuses are often not made public, generating fine-scale model-based estimates of settlement locations and population density has been an areas of substantial research focus for decades. 

The traditional approach to generating local-level population estimates takes a ``top-down" approach in which available admin-level census data is redistributed down to a finer-scale grid (1km or finer), using satellite-derived information and other covariates as input. Because population data are almost never available for training or validation at the target fine scale, one common approach uses the coarse-scale data from census to model the relationship between satellite features (e.g. nighttime lights imagery or satellite-derived estimates of land use), other ancillary data such as the location of transportation infrastructure, and census-based population estimates, and then applies the trained model to available fine-scale features\cite{dobson2000landscan,stevens2015disaggregating}.  Another approach generates a binary population mask at fine scale using estimates of building or settlement locations derived from imagery, and then applies this mask to coarse-scale census data\cite{schiavina2019ghs}. Both approaches typically use machine learning at some step, e.g a random forest to predict coarse census data, or computer vision approaches to identify settlement locations. For either approach, predictions can only be readily evaluated at coarse scale; the fine scale gridded predictions cannot be easily validated. In the absence of clear evaluation opportunities, a consortium of data producers have built useful tools in which different gridded estimates can be visually compared at local scale (https://popgrid.org).

\begin{figure}[H]
\begin{center}
  \includegraphics[width=\textwidth]{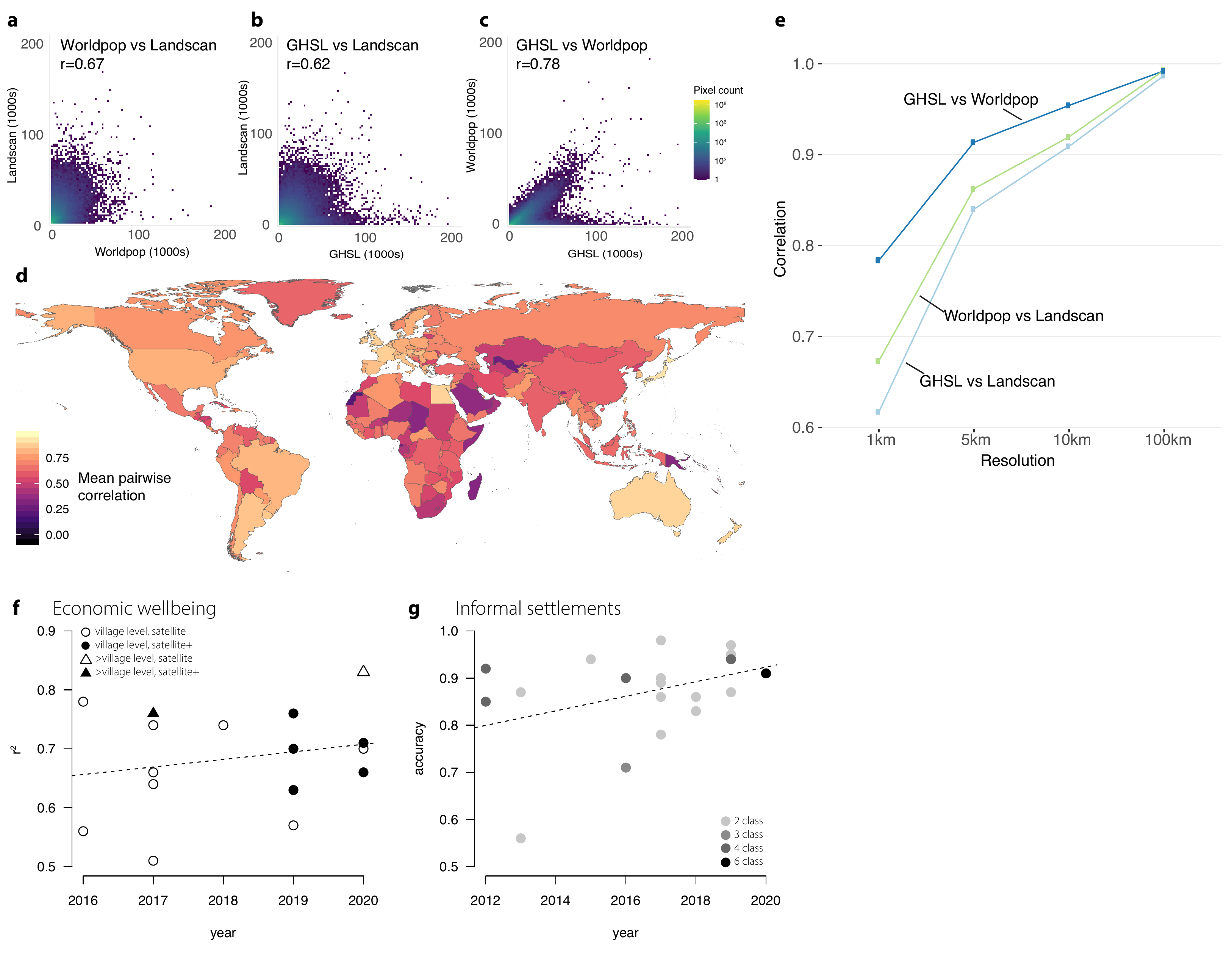}
\end{center}
\caption{\textbf{Performance of satellite-based approaches to measuring population, wealth, and informal settlements}. \textbf{a-c} Comparison of three different satellite-informed global population datasets (Landscan, WorldPop, and GHSL population) datasets at 1km resolution globally (colors correspond to scale at right). \textbf{d} Average pairwise correlation within each country at 1km resolution. Comparisons show modest correlation between datasets at global scale and often poor correlation in many developing countries. \textbf{e} Correlations across datasets improve when data are spatially aggregated. All comparisons are made for pixels that were not missing and not zero across all three datasets. \textbf{f} Performance in predicting asset wealth in various developing countries from satellite data (16 estimates from 12 papers), as measured by coefficient of determination on test data. Filled markers are estimates that combine satellite information with other data (cell phone data, social media data, or Wikipedia). Circles indicate estimates at the village level, triangles are estimates at more aggregate spatial scale (sub-district or district). \textbf{g} Performance in predicting the location of informal settlements from imagery (20 estimates from 17 papers}.
\label{fig:population}
\end{figure}

As additional quantitative comparison, we study three commonly-used population rasters that used satellite data as at least one input in their production: WorldPop\cite{stevens2015disaggregating}, GHSL\cite{schiavina2019ghs}, and LandScan\cite{dobson2000landscan}. We harmonize each to a consistent 1km grid and compare population estimates for grid cells with non-zero estimates across all three rasters. Estimates show modest agreement (r=0.62-0.78) when comparing across all global pixels (Figure \ref{fig:population}), with lowest agreement between LandScan and the other rasters. Agreement was often substantially lower in the developing world. On the African continent, the average pairwise correlation between the 3 datasets across 47 African countries is $r=0.45$, perhaps in part due to the relative paucity of census data on which to train models. Overall disagreements in African and globally could also result from differences in conceptualization of population used in dataset construction, with LandScan attempting to measure ``ambient population" averaged over 24 hours and the other datasets attempting to measure population at individuals' usual residences. Agreement improves when comparisons are made at increasingly aggregate levels, with correlations approaching $r=1.0$ when estimates are aggregated to 100km pixels.

Multiple studies have sought to further validate estimates of one or more of these datasets in settings where fine-scale population data are available. Using very-fine-scale (100m) administrative population data from Sweden available over a 25-yr period (none of which used in the creation of any of the gridded datasets), researchers found cell-wise correlations between the admin data and GHSL, WorldPop, and LandScan of $r=0.83$, $r=0.82$, and $r=0.7$, respectively, with predictive performance improving slightly in later years\cite{bustos2020pixel}. The authors caution that performance in Sweden (where model predictions were highly correlated, see Fig \ref{fig:population}) might not reflect performance elsewhere, given the high quality of ancillary data available in Sweden. Other studies in China and Europe found similar or higher performance of individual gridded datasets evaluated at somewhat more aggregate scales, but (as in Fig \ref{fig:population}) found that performance was not uniform and tended to degrade at finer spatial scales.\cite{calka2020ghs,bai2018accuracy} Overall performance on this population prediction task appears roughly on par with performance predicting asset wealth described below. 

Because standard approaches to generating these estimates is to disaggregate official census estimates, final estimates are unavoidably affected by any inaccuracies in the official census data -- for instance due to the most recent census having occurred a decade or more prior. %
An alternative that does not face this problem is to train ``bottom-up" models to directly predict local-level population estimates, and these approaches have shown promise in multiple settings\cite{wardrop2018spatially,engstrom2020estimating,hu2019mapping}. Such approaches are beginning to be incorporated into global gridded products (e.g. WorldPop) for countries where censuses are particularly out of date\cite{sdsn2020leaving}, and have been shown to be a cost-effective way for generating reliable national-scale population estimates\cite{wardrop2018spatially}.

\textbf{Economic livelihoods}

Predicting variation in local-level economic outcomes is another domain where the combination of machine learning and satellite imagery has seen recent application, again motivated by the paucity of existing data (Fig \ref{fig:surveys}) and the broad range of applications for which such data could be useful. As in the agricultural setting, existing work spans diverse geographies and seeks to predict a range of outcomes, making quantitative comparison of different models or sensors difficult.  

We focus on 12 studies that used imagery -- either alone or in combination with other data -- to predict asset wealth at local level in the developing world. Asset wealth is a commonly used measure of households' longer-run economic wellbeing, and is consistently measured in a number of georeferenced nationally-representative household surveys, making it appealing training data in this domain. Fig \ref{fig:population}a shows 16 asset wealth estimates across these 11 studies. All studies applied convolutional neural networks to imagery to generate features used to predict wealth, and reported evaluation statistics on held-out test data. 

While study intercomparison was challenging even for this group of studies that measured the same outcome due to the varied geographic settings (spanning Africa, Asia, and the Caribbean), the various spatial scales at which predictions were evaluated (from village level to district level), and some studies' inclusion of additional data input data not from satellites, results allowed some generalizations. First, information derived from satellites could always explain more than half, and often more than 75\%, of the variation in the survey-measured asset wealth, with performance appearing to trend upward over time. For reasons described above, these estimates likely understate true model performance, as test data are almost always from publicly-available survey data with known sources of noise.  Second, although small samples make generalization tenuous, studies that made predictions at more aggregate spatial scales, and studies that combined satellite information with data from other sources, tended to outperform village-level satellite-only models. These data fusion approaches have become increasingly common, with researchers demonstrating how combining imagery with data from cell phones\cite{steele2017mapping}, Wikipedia\cite{sheehan2019predicting}, social media\cite{fatehkia2020relative}, or Open Street Map\cite{tingzon2019mapping} can improve predictions. 

Table \ref{tab:poverty} describes results from additional studies that looked at other measures of economic livelihoods, including consumption expenditure and multi-dimensional poverty indices. Prediction performance for consumption expenditure (the measure on which official poverty estimates are based) is typically lower than that for asset wealth, a difference which has been in part attributed to relatively higher noise in the consumption data\cite{jean2016combining,yeh2020using} and the extreme paucity of public georeferenced public data on which to train models.

\textbf{Informal settlements} A final related area where there has been much recent work is in the detection of informal settlements (sometimes called “slums"). Urban populations are growing rapidly throughout much of the developing world, and about 30\% of developing-country urban populations are estimated to live in slums -- settled areas where inhabitants lack access to essential services, durable housing, and/or tenure security\cite{habitat2015habitat}. Systematic data on the location and size of such settlements is lacking, making it difficult to monitor and target service delivery and to protect residents against eviction, among other challenges\cite{kuffer2016slums}.  Some governments, lacking reliable data on informal settlements, do not officially acknowledge their existence\cite{habitat2015habitat}. 

Because the spatial structure (e.g density, size and type of buildings) can differ substantially between informal settlements and surrounding regions, researchers have sought to use imagery to measure the location and size of these settlements (see ref\cite{kuffer2016slums} for a recent review). We focus on 23 studies that used satellite imagery to segment or classify informal settlements in the developing world. These studies use a variety of methods, with some focused on creating rule bases for classification and others on directly using machine learning for classification. The fuzzy-logic rule bases are sometimes generated using machine learning (eg. decision trees) and sometimes are human generated from ontologies (formalized descriptions of expert knowledge from a certain perspective) of local informal settlements.

As with the other domains discussed, the literature spans diverse geogrophies where informal settlements can be very structurally dissimilar from each other, making study intercomparison difficult. However, in 17 studies that reported classification accuracy (evaluated against typically small numbers of ground observations), accuracy exceeded 80\% in most studies and appeared to be improving over time (Fig \ref{fig:population}g). 
Table \ref{tab:slums} shows results from additional studies that reported alternate performance metrics.

\subsection{Application in research}
Here we highlight a number of settings in which measures derived from satellite-based remote sensing, including those discussed above, are being used for some downstream research task in the developing world. 

The widest adoption of satellite-derived measures in research and policy has been in the realm of population estimates, with existing gridded population data being used in an impressive array of research applications. These include in public health, disaster response, economic development, climate change research, and others; see refs\cite{wardrop2018spatially,leyk2019spatial,sdsn2020leaving} for excellent recent reviews.

Satellite imagery has also been widely used to better understand agricultural productivity, including why some fields or some regions are more productive than others\cite{lobell2013use} and whether particular management practices have been adopted\cite{kubitza2020estimating}. Satellite estimates are also increasingly being used to identify fields most likely to respond to a particular input\cite{burke2017satellite,lobell2020eyes} or new management practice\cite{jain2019impact}.

Fisheries and animal production are additional food-related domains where satellite imagery is becoming increasingly used in research and policy. Recent work shows how multiple satellite sensors and deep learning can shed light on overall patterns of global fishing activity\cite{kroodsma2018tracking} as well as on specific activities like illegal fishing\cite{park2020illuminating,belhabib2020catching}. %

Researchers in economics also increasingly utilize satellite imagery -- and particularly night-time lights imagery -- for a variety of applications (see ref\cite{donaldson2016view} for a review).  Nightlights have been used to assess the validity of official government statistics\cite{henderson2012measuring,pinkovskiy2016lights}, to understand the growth and activity of urban versus rural areas\cite{henderson2018global,harari2020cities}, and to assess the role of local and federal institutions, transport costs, and other factors on economic development\cite{michalopoulos2013pre,michalopoulos2014national,storeygard2016farther,pinkovskiy2017growth}. While the use of optical imagery beyond nightlights remains somewhat more limited, recent papers have shown how high-resolution optical imagery can be used to measure compliance with conservation programs\cite{jayachandran2017cash} and to understand how ethnic favoritism shapes economic investment\cite{marx2019there}.
Recent work~\cite{zhao2020framework} also shows how to combine satellite-derived estimates with survey data to obtain tighter confidence intervals and improve regression analyses.

More recent work has shown how satellites can be useful in the experimental evaluation of interventions in both the agricultural and economic sphere. Jain et al\cite{jain2019impact} show how remote sensing estimates can be used to measure the effectiveness of a new agricultural technology on productivity and quantify who benefits most from the adoption of the technology. Huang\cite{huang2020measuring} shows how a deep learning model trained to identify housing quality in high-resolution imagery can be used to estimate the livelihood impact of a randomized cash transfer program in Kenya, with estimates benchmarked against ground survey data.  Jayachandran et al\cite{jayachandran2017cash} show how high-resolution imagery can be used to measure compliance in an experimental evaluation of a payment-for-forest-protection program. While all of these studies focus on settings where changes induced by an intervention are readily apparent in imagery -- an aspect that might not hold in other settings -- they demonstrate the large potential for satellite imagery to contribute to the quantitative evaluation of many development interventions.

\subsection{Use in decision making}

While satellite-based measures are now being used in a variety of research applications, documented examples of their operational use in public-sector decision-making and policy in the developing world is much more limited. Systematic information on operational use in the private sector is even more sparse, although use is likely widespread and growing; the same is true of military applications. Here we only consider public-sector non-military use. 

As in research, the widest application of satellite-based measures in public-sector decision-making is in the population domain. For instance, the UN World Food Programme and US government both used gridded population estimates to inform needs assessments and target humanitarian response following natural disasters\cite{sdsn2020leaving}. Gridded population data are also being used to inform sampling strategies for ground surveys\cite{sdsn2020leaving}.

In agriculture, remote-sensed vegetation indices and satellite-derived rainfall estimates are key inputs into short-term forecasting of food insecurity, which directly informs food aid and other humanitarian resource allocation\cite{brown2008famine}. Numerous systems that track agricultural growing conditions and crop output around the world also make ample use of remote sensing information, and output from these systems are used in a wide array of tasks, including in early warning alerts, foreign aid decisions, analysis of commercial trends, and in trade policy\cite{fritz2019comparison}. Data from remote detection of fishing activity is also being used by numerous governments and other organizations to manage fisheries and design protected areas\cite{gfw2020}.

Across other domains -- e.g. economic livelihood measurement -- documented use in decision-making appears limited or non-existent, although anecdotally there is rapidly growing interest in the policy community in exploring these measures\cite{blumenstock2020machine}.

We hypothesize on why adoption in these and other domains has been relatively limited. The simplest explanation is that the combination of satellite information and machine learning is still quite new in many domains, and decision-makers might not be familiar with these approaches or convinced they are ``good enough". Our view is that in many settings, including smallholder agricultural and livelihood measurement, the true accuracy of satellite-derived estimates can rival or exceed that of traditional survey-based measures. It remains the job of the research community to help make this clear, and the job of the user community to transparently define the counterfactual:  if not satellite-based data, what alternative data would be used to make a decision, and what do we know about its reliability?

Even if satellite-based measures are accurate, they might not yet be operational. To our knowledge there exist no updated, global-scale estimates of smallholder crop productivity, economic well-being, or informal settlements that a decision-maker could immediately use (estimates are beginning to exist for individual countries). The research community is arguably not well positioned to generate and update such estimates over time, and partnerships with public-sector institutions or the private sector to scale and operationalize these estimates could be important in enabling their sustained use.

Even when models are operational, decision-makers might be understandably hesitant to adopt a measure they cannot fully explain. Deep learning models tend to sacrifice interpretability for predictive performance, and researchers are often satisfied if a model is working well (as evaluated on held-out data) even if they cannot explain why. But understanding why a model makes the predictions it does can help build trust that predictions are accurate and fair. Well-publicized instances of algorithmic bias in other settings (e.g. predictive policing, sentencing, and hiring decisions\cite{cossins2018discriminating}), and concerns by civil rights groups that further deployment of algorithmic decision-making might worsen racial and socioeconomic inequalities\cite{d4bl,civilrights_data}, understandably amplify worries that predictions from these new approaches could be either inaccurate or unfair. 

Existing guidelines for Fairness, Accountability, and Transparency in Machine Learning (``FAT ML")\cite{diakopoulos2017principles}, if followed, could help navigate these issues. The guidelines aim to ensure that researchers are aware of potential discriminatory impacts of their algorithms and are able to investigate and provide redress should issues arise. While implementation of the guidelines certainly has its own challenges\cite{gajane2017formalizing} (e.g. defining ``fairness"), we are not aware of any of the papers we review above -- including our own -- having fully engaged with these guidelines.

A final reason for limited adoption is that some actors might see benefit in not having certain outcomes be measured. Autocratic regimes already collect less data (recall Fig \ref{fig:surveys}), and certain countries have passed laws (since reversed) that make it a crime to publish independent estimates of key economic outcomes.\cite{thecitizen_2019}

\section{Conclusions and directions for future work}

We draw four main conclusions from the above analysis, and lay out open challenges and directions for future work. First, satellite-based performance in predicting key sustainable development outcomes is reasonably strong and appears to be improving. Estimates are being used in a wide variety of research applications and, in some cases, are already actively informing decision-making. Indeed, analyses suggest that reported model performance likely understates true performance in many settings, given the noisy data on which predictions are evaluated, and that satellite-based estimates can equal or exceed the accuracy of traditional approaches to measuring key outcomes. For certain outcomes, satellite-based approaches can already add substantial information at broad scale and low cost compared to what can be collected on the ground. Numerous quantitative approaches now exist to assist researchers and practitioners in better understanding -- and not underestimating -- the performance of satellite-based approaches relative to traditional alternatives. 

Second, perhaps the largest constraint to model development is now training data rather than imagery. While imagery has become abundant, the scarcity and (in many settings) unreliability of quality labels make both training and validation of satellite-based models difficult. Expanding the quantity and -- in particular -- the quality of labels will quickly accelerate progress in this field, and allow both researchers and practitioners to measure new outcomes and to accurately assess model performance. 

Third, despite the growing power of satellite-based approaches, there are many domains where such approaches are likely to contribute little in the near term -- for instance, in measuring female empowerment, educational outcomes, or conflict events.  Even in settings where satellites are likely to be useful, satellite-based approaches will likely amplify rather than replace existing ground-based data collection efforts. 
High-quality local training data can nearly always improve model performance, and will remain essential for convincing both researchers and decision-makers that satellite-based approaches are working.

Finally, there remain limited documented cases where satellites have been operationalized into decision-making processes in the sustainable development domains where we focus -- with satellite-informed population estimates being the main exception. Limited adoption is likely driven by a number of forces, including the recency of the technology, the lack of accuracy (perceived or real) of the models, lack of model interpretability, and entrenched interests in maintaining the current data regime.  

Helping to overcome these constraints constitute key tasks for researchers and policymaker going forward. We suggest nine specific areas where we believe future work would be particularly useful:

\ben
    \item \textit{More accurate, more numerous training data}. Many applications of deep learning outside sustainable development have been advanced by the curation of reference datasets that are then made available to the community. These datasets lower the barriers to entry and make comparison of different approaches more straightforward, yet they are lacking for sustainable development outcomes. Particularly needed are datasets that track outcomes over time so that models can be optimized to detect changes.  These datasets are a major public good and should be funded as such.  Collecting and publishing location data from existing and ongoing ground surveys (using appropriate privacy safeguards already widely in use) should be mandated by survey funders.  
    \item \textit{More evaluation in the context of specific use cases}. Most evaluation of satellite estimates have focused on agreement with a ground-based measure of a particular outcome. Fewer studies have then gone the next step to evaluate the actual application of the outcome measure, such as to test the impact of a randomized control trial or target an intervention to a sub-population. These downstream tasks often provide a more tangible example of the utility to potential users, and can avoid the pitfalls of direct comparisons to noisy ground measures. A related task will be to define and utilize meaningful loss functions for the specific task at hand; for instance, a poverty targeting application might be more tolerant of small errors at the wealthy end of the distribution than the poorer end.
    \item \textit{Improved model interpretability and transparency}. Especially in cases where satellite-based prediction is being used to make decisions that directly impact people (e.g. targeting aid) it is especially important that predictions be explainable and that decisions based on those predictions be transparent. Applying FAT ML or similar guidelines to research output will be increasingly important as research gets operationalized.
    \item \textit{Creative data fusion}. Combining information from multiple different optical sensors of different temporal and spatial resolutions,  combining different types of imagery (e.g. optical + radar), and/or combining satellite imagery with other relevant data (e.g. from cell phones), appear to be particularly promising approaches to improving model performance. As much of these additional data are collected by the private sector, sustained and enforceable data-sharing agreements between companies and researchers will be key\cite{lazer2020computational}.
    \item \textit{Scaling estimates}. Researchers typically have more incentive to innovate on methods than they do (e.g.) to apply validated methods across large geographies and update estimates as new data come in -- the later being what is often required to make outputs useful to decision-makers. Partnerships between academic researchers and public- or private-sector organizations who have the skills and resources to do this scaling will be key to operationalizing many promising research advances in the satellite/ML domain.
    \item \textit{Measuring changes over time}. Much of the literature reviewed above makes predictions at a given point in time. However, many applications require measuring changes over time. While the relationship between inputs and outputs over time is reasonably stable in some domains (e.g. vegetation indices and yields in agriculture), this might not be true in other domains (e.g. economic development).  Unfortunately, temporal evaluation at a local level is difficult because there exist few ground datasets that repeatedly and reliably measure the same locations over time.  Curating these datasets and using them to develop and validate temporal predictions will be key for tracking the evolution of key sustainability outcomes. 
    \item \textit{Using imagery to actively guide ground data collection}. As predictive performance of satellite-based models improve, their output could be used to optimally guide further data collection on the ground -- for instance, to collect data in locations where model predictions are least certain. Research should explore to what extent such sampling strategies could improve outcome measurement compared to traditional sampling approaches. 
    \item \textit{Understanding potential pitfalls in causal inference applications}. For instance, can poverty predictions from a satellite-based model be used to study the impact of new road construction on poverty, if there is a chance that the model looks for a road to decide whether a location is poor? How do we proceed if we're concerned that image-derived proxies for a dependent variable of interest are themselves the independent variable of interest?  
    \item \textit{Improved guidelines for privacy}. As predictions become increasingly granular and accurate, who has access to these data?  How can precisely georeferenced ground data (which is increasingly collected) be used to train or validate models without undermining privacy? Guidelines for navigating these issues are increasingly critical as models improve. 
\een

\bibliographystyle{naturemag.bst}
\footnotesize{\bibliography{refs.bib}}

\newpage
\clearpage

\newpage
\clearpage

\newpage
\clearpage
\thispagestyle{empty}
\begin{figure}
\begin{center}
   \includegraphics[width=0.9\textwidth]{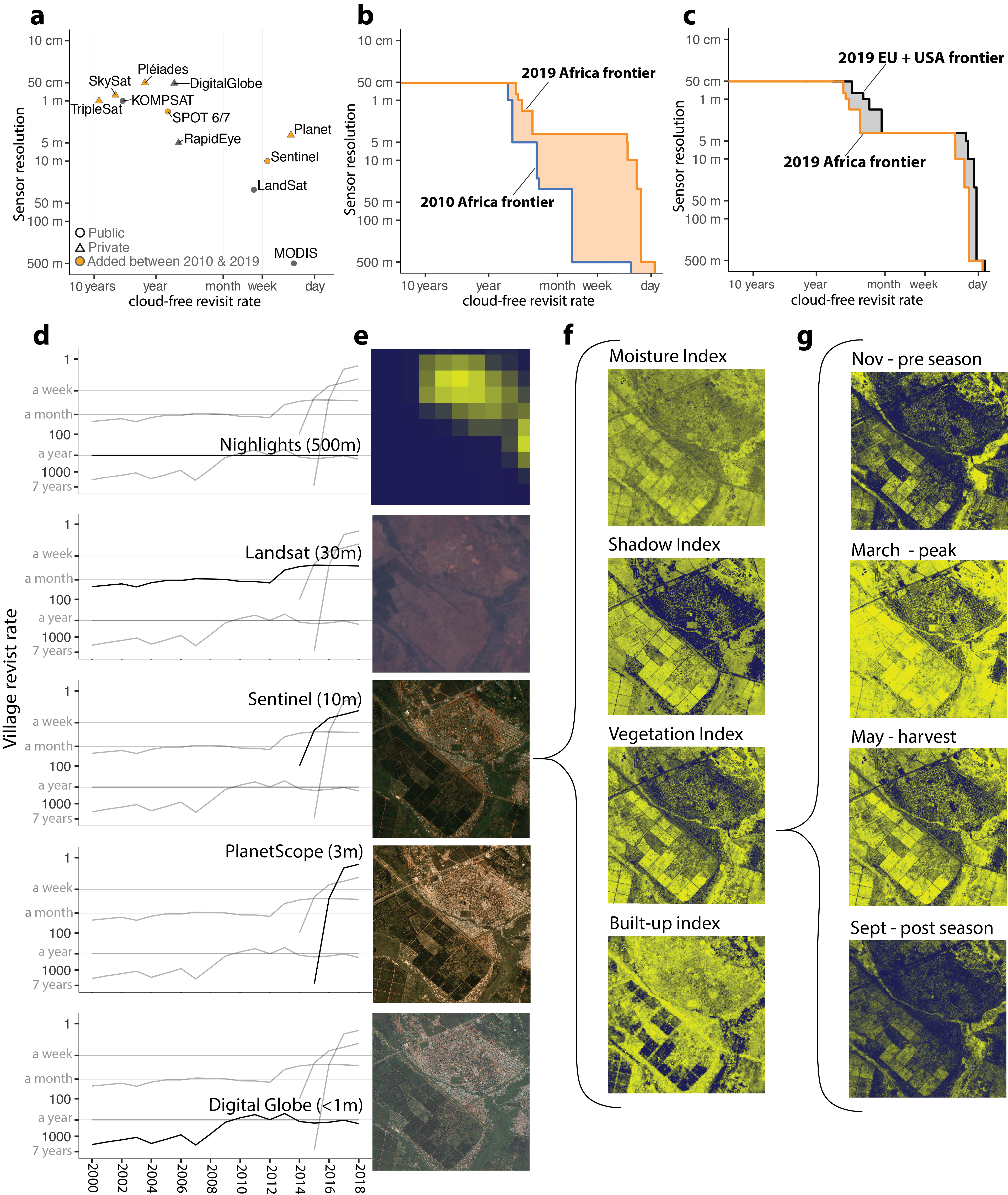}
\end{center}
\caption{\footnotesize{\textbf{Spatial resolution, temporal frequency, and spectral availability of satellite imagery have increased substantially since 2000}. \textbf{a} Average revisit rate and sensor resolution of cloud-free optical imagery in 2019, averaged across 100 populated African locations. \textbf{b} Blue line (``frontier") shows overall revisit rate across all available sensors at a given spatial resolution in 2010 for same 100 locations (e.g. at 1m, line denotes average cloud-free revisit rate using all sensors $<=$1m); orange line shows same for 2019. Orange area denotes the new combinations of temporal and spatial resolution available by 2019, which expanded greatly at resolutions $>$1m \textbf{c} Average 2019 coverage in Africa (orange line) vs 100 locations in US/EU (grey line; locations randomly sampled proportional to population). Grey shaded area depicts inequalities in coverage between wealthy and developing regions in 2019, which are larger for imagery $<$3m/px. \textbf{d} Calculated revisit periods for several satellites over 500 randomly selected survey locations in Africa since 2000. Nightlights is set to a one year revisit rate given the stable yearly product. \textbf{e} Example imagery corresponding to each sensor in a single location in central Zambia. Images are real color except for NL. \textbf{f} Indices generated from various bands can convey different information, as depicted here using Sentinel 2 data (yellow colors indicate higher values of the index).  \textbf{g} Frequent revisit rates of new public sensors capture temporal variation in human activity, including rapid changes throughout the main agricultural season shown here.}}
\label{fig:imagery}
\end{figure}

\newpage
\clearpage
\thispagestyle{empty}

\newpage
\clearpage

\newpage
\clearpage

\newpage
\clearpage

\normalsize
\renewcommand\thefigure{A\arabic{figure}}
\renewcommand\thetable{A\arabic{table}}
\setcounter{figure}{0}
\setcounter{table}{0}

\section*{Supplementary Information}
\renewcommand\thefigure{S\arabic{figure}}
\renewcommand\thetable{S\arabic{table}}
\setcounter{figure}{0}
\setcounter{table}{0}

\subsection*{Collecting Satellite Revisit Data}

Construction of Figure \ref{fig:imagery} involved acquiring data from several sources. First we use Gridded Population of the World (GPW) data raster data to create a population weighted sample of 100 locations in Africa, as well as 100 locations across the EU and the USA. These 200 locations are then buffered with an approximately 10 meter radius and used to query for satellite imagery for the years 2010 and 2019. 

Planet products (SkySat, PlanetScope and RapidEye) were downloaded from the Planet API \cite{planetapi}. Footprints for other private satellites were downloaded from LandInfo \cite{landinfo}, while footprints from public satellites (Landsat, Sentinel, MODIS) were downloaded using Google Earth Engine \cite{gorelick2017google}. 

We attempted to maintain consistency in filtering across data sources, but filtering works slightly differently in each system. LandInfo (though it lacks in depth documentation confirming this) appears to filter solely over the area of interest (AOI) rather than over the entire footprint. Public data was processed to match this, using only the cloud cover percentage over the buffered polygon, but Planet is filtered at the footprint level. All image footprints were filtered to be <30\% cloud cover and off-nadir <|20|.

Sensors in Figure \ref{fig:imagery}a are grouped slightly to make the figure easier to process visually. Landsat 7 and 8 are combined; WorldView-1 through 4, GeoEye-1, QuickBird-2, and IKONOS are grouped as ``DigitalGlobe"; KOMPSAT-3, KOMPSAT-3A, and KOMPSAT-2 are grouped; SPOT-4 and 5 are grouped as well as SPOT-6 and 7. As sensor resolution varies within these groups, we use the mode of the resolutions in the group to represent the group as a whole. This does compress the range of resolutions significantly, for example ``DigitalGlobe" is recorded as a resolution of 51cm, where the true resolutions range from 31cm to 91cm. ``KOMPSAT" ranges from 70cm to 1m, ``SkySat" ranges from 70cm to 1m, ``SPOT 4/5" ranges from 2.5m to 10m. 

To calculate the average revisit rate, we sum up the total number of images collected in each group and calculate (number of locations*365)/number of images. For the frontier, we calculate the revisit rate by summing the total number of images collected for all satellites with resolution less than or equal to the resolution of interest and run the same calculation as above. As this is an average of time between images, a number below 1 does not necessarily indicate that there is a cloudless picture on every day.

\newpage
\clearpage
\begin{table}[]
    \centering
    \caption{Performance of studies using satellites to predict smallholder yields at the plot level. Year=study year, res=sensor resolution, n=number of observations, r2=squared correlation, data=training data (CC=crop cuts, SR=self reported, FP=full plot). In the sensor column, ``Skysat (c)" refers to Skysat data that has been coarsened to lower resolution.}
    \begin{tabular}{rlrlllrrrl}
  \hline
 & study & year & location & crop & sensor & res & n & r2 & data \\ 
  \hline
1 & Jain et al 2016\cite{jain2016mapping} & 2014 & India & wheat & Skysat & 2 & 50 & 0.27 & CC \\ 
  2 & Jain et al 2016\cite{jain2016mapping} & 2015 & India & wheat & Skysat & 2 & 37 & 0.33 & CC \\ 
  3 & Jain et al 2016\cite{jain2016mapping} & 2014 & India & wheat & Skysat & 2 & 52 & 0.13 & SR \\ 
  4 & Jain et al 2016\cite{jain2016mapping} & 2015 & India & wheat & Skysat & 2 & 29 & 0.06 & SR \\ 
  5 & Jain et al 2016\cite{jain2016mapping} & 2014 & India & wheat & Landsat & 30 & 50 & 0.52 & CC \\ 
  6 & Jain et al 2016\cite{jain2016mapping} & 2015 & India & wheat & Landsat & 30 & 37 & 0.42 & CC \\ 
  7 & Lambert et al 2017\cite{lambert2017estimate} & 2016 & Mali & cotton & S2 & 10 & 9 & 0.64 & CC \\ 
  8 & Lambert et al 2017\cite{lambert2017estimate} & 2016 & Mali & maize & S2 & 10 & 9 & 0.58 & CC \\ 
  9 & Lambert et al 2017\cite{lambert2017estimate} & 2016 & Mali & millet & S2 & 10 & 8 & 0.84 & CC \\ 
  10 & Lambert et al 2017\cite{lambert2017estimate} & 2016 & Mali & sorghum & S2 & 10 & 9 & 0.41 & CC \\ 
  11 & Guan et al 2017\cite{guan2018mapping} & 2015 & Vietnam & rice & Landsat & 30 & 71 & 0.40 & CC \\ 
  12 & Karst et al 2020\cite{karst2020estimating} & 2018 & Burkina Faso & beans & S2 & 10 & 31 & 0.50 & CC \\ 
  13 & Karst et al 2020\cite{karst2020estimating} & 2018 & Burkina Faso & maize & S2 & 10 & 31 & 0.40 & CC \\ 
  14 & Karst et al 2020\cite{karst2020estimating} & 2018 & Burkina Faso & sorghum & S2 & 10 & 57 & 0.40 & CC \\ 
  15 & Karst et al 2020\cite{karst2020estimating} & 2018 & Burkina Faso & millet & S2 & 10 & 45 & 0.32 & CC \\ 
  16 & Jin et al 2017\cite{jin2017mapping} & 2016 & Kenya & maize & S2 & 10 & 41 & 0.36 & CC \\ 
  17 & Schulthess et al 2013\cite{schulthess2013mapping} & 2010 & Bangladesh & maize & rapideye & 5 & 30 & 0.56 & SR \\ 
  18 & Zhao et al 2017\cite{zhao2017detecting} & 2009 & China & rice & formosat-2 & 8 & 22 & 0.11 & SR \\ 
  19 & Lobell et al 2020\cite{lobell2020sight} & 2017 & Mali & sorghum & S2 & 10 & 575 & 0.23 & CC \\ 
  20 & Lobell et al 2020\cite{lobell2020sight} & 2017 & Mali & sorghum & Planet & 3 & 575 & 0.25 & CC \\ 
  21 & Lobell et al 2020\cite{lobell2020sight} & 2017 & Mali & sorghum & S2 & 10 & 575 & 0.05 & SR \\ 
  22 & Lobell et al 2020\cite{lobell2020sight} & 2017 & Mali & sorghum & Planet & 3 & 575 & 0.06 & SR \\ 
  23 & Burke and Lobell 2017\cite{burke2017satellite} & 2014 & Kenya & maize & Skysat & 1 & 72 & 0.39 & SR \\ 
  24 & Burke and Lobell 2017\cite{burke2017satellite} & 2014 & Kenya & maize & Skysat (c) & 5 & 72 & 0.38 & SR \\ 
  25 & Burke and Lobell 2017\cite{burke2017satellite} & 2014 & Kenya & maize & Skysat (c) & 10 & 72 & 0.30 & SR \\ 
  26 & Burke and Lobell 2017\cite{burke2017satellite} & 2014 & Kenya & maize & Skysat (c) & 30 & 72 & 0.25 & SR \\ 
  27 & Burke and Lobell 2017\cite{burke2017satellite} & 2014 & Kenya & maize & Skysat & 1 & 386 & 0.20 & SR \\ 
  28 & Burke and Lobell 2017\cite{burke2017satellite} & 2014 & Kenya & maize & Skysat (c) & 5 & 386 & 0.18 & SR \\ 
  29 & Burke and Lobell 2017\cite{burke2017satellite} & 2014 & Kenya & maize & Skysat (c) & 10 & 386 & 0.15 & SR \\ 
  30 & Burke and Lobell 2017\cite{burke2017satellite} & 2014 & Kenya & maize & Skysat (c) & 30 & 386 & 0.10 & SR \\ 
  31 & Lobell et al 2019\cite{lobell2020eyes} & 2016 & Uganda & maize & S2 & 10 & 252 & 0.55 & FP \\ 
  32 & Lobell et al 2019\cite{lobell2020eyes} & 2016 & Uganda & maize & S2 & 10 & 252 & 0.25 & SR \\ 
  33 & Lobell et al 2019\cite{lobell2020eyes} & 2016 & Uganda & maize & S2 & 10 & 252 & 0.25 & CC \\ 
   \hline
\end{tabular}
    \label{tab:ag}
\end{table}

\newpage
\clearpage
\begin{table}[]
    \centering
    \caption{Performance of efforts to predict economic wellbeing with imagery and ML. ``Samples" reports the number of training examples when available. ``Geo" reports the geographic level at which models were evaluated, with (e.g.) "geo2" equal to county or district.}
    
\begin{center}
 \begin{tabular}{||M{7pt} |M{23pt} |M{95pt} | C{40pt} |C{31pt} | C{80pt} | M{50pt} | M{55pt} ||} 

 \hline
 & Year & Target & Metric & Result & Samples & Geo & Location \\ [0.5ex] 
 \hline \hline

\cite{yeh2020using} & 2020 & Asset Wealth Index & r2 & 0.83 & 19,669 clusters & geo2 & 23 African countries \\
 \hline
\cite{kim2016incorporating} & 2016 & Asset Wealth Index & r2 & 0.78 & & cluster & Nigeria\\
 \hline
\cite{steele2017mapping} & 2017 & Wealth Index & r2 & 0.76 & & voroni polygons & Bangladesh \\
 \hline
\cite{jean2016combining} & 2016 & Asset Wealth Index & r2 & 0.75 & 1,411 clusters & cluster & Rwanda \\
 \hline
\cite{perez2017poverty} & 2017 & Asset Wealth Index & r2 & 0.66 & & cluster & Africa\\
 \hline
\cite{engstrom2017poverty} & 2017 & Income <= National 10th percentile & r2 & 0.61 & 1,291 villages & geo4 & Sri Lanka\\
 \hline
\cite{ayush2020generating} & 2020 & Consumption & r2 & 0.54 & 320 clusters & cluster & Uganda\\
 \hline
 \end{tabular}
 
\begin{tabular}{||M{7pt} |M{23pt} |M{95pt} | C{40pt} |C{30pt} | C{80pt} | M{50pt} | M{55pt} ||} 
 \hline
\cite{pokhriyal2017combining} & 2017 & Multi-Dimensional Poverty Index & cor & 0.91 & 552 communes & geo4 & Senegal\\
 \hline
\cite{njuguna2017constructing} & 2016 & Multi-Dimensional Poverty Index & cor & 0.88 & 416 sectors & geo3 & Rwanda\\
 \hline
\cite{perez2019semi} & 2019 & Asset Wealth Index & cor & 0.57 & 4,839 clusters & cluster & Africa \\ 
 \hline
 \end{tabular}
 
\begin{tabular}{||M{7pt} |M{23pt} |M{95pt} | C{40pt} |C{30pt} | C{80pt} | M{50pt} | M{55pt} ||} 
 \hline
\cite{smith2018left} & 2017 & If below Comparative Wealth Index poverty line or not & accuracy & 0.83 & eval on 636,448 hholds & cluster & 36 countries\\
 \hline
\cite{irvine2017viewing} & 2017 & Reported living condition good/neutral/bad & accuracy & 0.83 & & cluster & Botswana, Kenya, Zimbabwe\\
 \hline
\cite{li2019comparison} & 2019 & If county is "non-poverty" or not & accuracy & 0.82 & 192 counties & geo2 & China \\
  \hline
\cite{xie2016transfer} & 2016 & Above or below poverty line & accuracy & 0.72 & 643 clusters & cluster & Uganda\\
  \hline
\cite{watmough2019socioecologically} & 2018 & Bottom 40\%, middle 40\%, top 20\% classification & accuracy & 0.62 & 330 hholds & hhold & Kenya\\
  \hline
\cite{watmough2016understanding} & 2016 & Welfare Index quintiles & accuracy & 0.36 & 14,000 clusters & cluster & India\\
  \hline
\end{tabular}
\end{center}

    \label{tab:poverty}
\end{table}

\newpage
\clearpage

\begin{table}[]
    \centering
    \caption{Performance of efforts to predict the location of informal settlements (slums) with imagery and ML}
    \begin{center}

 \begin{tabular}{|| M{7pt}|M{1cm}|M{5cm}|C{1.5cm}|C{31pt}|M{80pt}|M{65pt} ||} 
 \hline
  & Year & Target & Metric & Result &  Samples & Location \\ [0.5ex] 
 \hline \hline 
\cite{mahabir2020detecting} & 2018 & MajiData spatial extent of slums & recall & 0.95 & & Kenya \\
\hline
\cite{rhinane2011detecting} & 2011 & Slum delineations & recall & 0.85 & 70km$^2$ classified & Morocco \\
 \hline 
\cite{hofmann2008detecting} & 2008 & Manual slum delineations & recall & 0.68 & & Brazil \\
 \hline 
  \end{tabular}
 
 \begin{tabular}{|| M{7pt}|M{1cm}|M{5cm}|C{1.5cm}|C{31pt}|M{80pt}|M{65pt} ||} 
 \hline
\cite{stoler2012assessing} & 2012 & Slum index & r2 & 0.4 & eval on 1,724 EAs & Ghana \\
 \hline 
  \end{tabular}
  
 \begin{tabular}{|| M{7pt}|M{1cm}|M{5cm}|C{1.5cm}|C{31pt}|M{80pt}|M{65pt} ||} 
 \hline
\cite{maiya2018slum} & 2018 & Slum Delineations & IoU & 0.9 & & India \\
 \hline 
  \end{tabular}
 
 \begin{tabular}{|| M{7pt}|M{1cm}|M{5cm}|C{1.5cm}|C{31pt}|M{80pt}|M{65pt} ||} 
 \hline
\cite{duque2017exploring} & 2017 & Slum delineations & accuracy & 0.98 & 12,398 100m cells & Colombia\\
\hline
\cite{gram2019mapping} & 2019 & Annotated ground truth points for slums & accuracy & 0.97 & & India \\ 
\hline
\cite{engstrom2015mapping} & 2015 & Accra Metropolitan Assembly slum dichotomy map & accuracy & 0.94 & 3,000 samples & Ghana \\
\hline
\cite{williams2019mapping} & 2019 & Point locations of squatter settlements & accuracy & 0.94 & & Jamaica\\
\hline
\cite{verma2019transfer} & 2019 & Slum delineations & accuracy & 0.94 & & India\\
\hline
\cite{leonita2018machine} & 2018 & Slum delineations & accuracy & 0.94 & & Indonesia\\
\hline
\cite{graesser2012image} & 2012 & ? & accuracy & 0.92 & 12,000 points & Afghanistan \\
\hline
\cite{fallatah2020object} & 2020 & Slum delineations & accuracy & 0.91 & & Saudi Arabia \\
\hline
\cite{persello2017deep} & 2017 & Slum delineations & accuracy & 0.9 & 3,000 samples & Tanzania \\
\hline
\cite{mboga2017detection} & 2017 & Slum delineations & accuracy & 0.9 & 3,060 samples & Tanzania\\
\hline
\cite{jochem2018identifying} & 2018 & Regular/irregular settlement 1km grid & accuracy & 0.9 & & Afghanistan \\
\hline
\cite{wurm2017slum} & 2017 & Slum areas from visual image interpretation & accuracy & 0.89 &  1,159,662 pixels & India \\
 \hline
\cite{kuffer2016extraction} & 2016 & Municipality provided slum location data & accuracy & 0.88 & 80 points & India\\
 \hline
\cite{kit2013automated} & 2013 & Rule based classification, using 7 ground truth locations & accuracy & 0.87 & eval on 7 points & India\\
\hline
\cite{hofmann2017object} & 2017 & Manual classification map & accuracy & 0.78 & eval on region & South Africa\\
 \hline
\cite{kohli2016urban} & 2016 & Slum delineations & accuracy & 0.71 & eval on 250 pts & India\\
 \hline
\cite{kohli2013transferability} & 2013 & Manual slum delineations & accuracy &  0.56 & eval on city & India\\
 \hline

\end{tabular}
\end{center}

    \label{tab:slums}
\end{table}

\end{document}